\documentclass{PoS}
\usepackage{cite}
\usepackage{amsmath}
\usepackage{amssymb}
\usepackage{url}

\title{Curvature of the QCD critical line with 2+1 HISQ fermions}

\ShortTitle{Curvature of the QCD critical line with 2+1 HISQ fermions}

\author{\speaker{Leonardo Cosmai}%
\\
        INFN - Sezione di Bari, I-70126 Bari, Italy\\
        E-mail: \email{leonardo.cosmai@ba.infn.it}}
\author{Paolo Cea\\
            Dipartimento di Fisica dell'Universit\`a di Bari, I-70126 Bari, Italy \\ and INFN - Sezione di Bari, I-70126 Bari, Italy\\
            E-mail: \email{paolo.cea@ba.infn.it}}
\author{Alessandro  Papa\\
            Dipartimento di Fisica, Universit\`a della Calabria, \\
            and INFN - Gruppo Collegato di Cosenza, I-87036 Rende, Italy \\
            E-mail: \email{papa@cs.infn.it}}

\abstract{
We present results on the curvature of the critical line of QCD with 2+1 HISQ fermions at nonzero temperature and quark density obtained by analytic continuation from imaginary chemical potentials. Monte Carlo simulations are performed by means of the MILC code, suitably modified to include a nonzero imaginary baryon chemical potential. We set the chemical potential at the same value for the three quark species  and work on the line of constant physics with a light to strange mass ratio of 1/20 as determined in Ref.~\cite{Bazavov:2011nk}.}

\FullConference{The 32nd International Symposium on Lattice Field Theory\\
		 23-28 June, 2014\\
		 Columbia University New York, NY}

\begin{document}

\section{Introduction}
\label{introd}

Quantum Chromodynamics 
(QCD) is the theory underlying strong interactions and can be used 
to account for the different phases 
of strongly interacting  matter  under usual or unusual (extreme) conditions.
A transition or rapid crossover is thought to exist 
from a   low-temperature  hadronic phase to a   high-temperature  
Quark-Gluon Plasma (QGP) phase; 
the line separating these two phases in the
temperature - baryon density plane is called the QCD (pseudo)critical line.
The  location of this line and the nature of 
the transition across it has many important theoretical and
phenomenological implications, which
go from the physics of the early Universe (high $T$ - low baryon density region of the phase diagram), to the physics
of the interior of some compact astrophysical objects (corresponding
to the low $T$ - high density region).
Various experiments have been devised or have been planned
in order to study this transition via heavy-ion collisions at ultrarelativistic energies. 
Depending on the beam energy, different conditions of temperature and baryon 
density can be realized in the fireball produced 
after the collision, such that the QGP phase appears as
a transient state, before the system freezes out and partons recombine into
hadrons. For a given collision energy, the particle yields are found to
be well described by a thermal-statistical model assuming approximate 
chemical equilibrium, as realized at the chemical freeze-out
point, in terms of only two parameters, the freeze-out 
temperature $T$ and the baryon chemical potential $\mu_B$. The set of
freeze-out parameters determined in experiments with different collision
energies lie on a curve in the $(T,\mu_B)$-plane, extending up to
$\mu_B\lesssim$ 800 MeV (see Fig.~1 of Ref.~\cite{Cleymans:2005xv},
or Ref.~\cite{Becattini:2012xb} for a recent re-analysis of experimental data).
Chemical freeze-out is reached as the fireball
cools down, subsequently to re-hadronization. 
A reasonable guess is that chemical freeze-out 
is reached shortly after hadronization, so that the QCD (pseudo)critical line and the freeze-out 
curves lie close to each other. In general they can be parametrized, at low baryon densities, by
a lowest order Taylor expansion in the baryon chemical potential, as follows
\begin{equation}
\frac{T(\mu_B)}{T_c(0)}=1-\kappa \left(\frac{\mu_B}{T(\mu_B)}\right)^2\;,
\label{curv}
\end{equation}
where $T_c(0)$ is the (pseudo)critical temperature at vanishing baryon density.

The QCD (pseudo)critical line can be determined within a first-principle approach exploiting  lattice gauge theory methods. But at nonzero baryon density,
due to the well known "sign problem", the QCD fermion determinant becomes complex and standard numerical simulations are unfeasible. 
Several methods have been invented to attack this problem (for a review, 
see~\cite{Philipsen:2005mj,Schmidt:2006us,deForcrand:2010ys,Aarts:2013bla}).
In the present work we exploit the method of analytic continuation from an imaginary chemical potential to
give a first estimate of the QCD (pseudo)critical line using the HISQ/tree action~\cite{Follana:2006rc,Bazavov:2010ru} 
with 2+1 staggered fermions (see also Ref.~\cite{Cea:2014xva}).
The strange mass is set 
at the physical value and simulations are performed on the line of constant 
physics (LCP) with the light quark mass fixed at $m_l=m_s/20$, as determined 
in Ref.~\cite{Bazavov:2011nk}.  As for quark chemical potentials, in the
present study we assign the same value to the three quark species, 
$\mu_l=\mu_s\equiv \mu$. We explore lattices of different
spatial extensions, $16^3\times 6$ and $24^3\times 6$, 
to check for finite size effects, and present 
results on a $32^3 \times 8$ lattice,
to check for finite cut-off effects.

\section{Numerical results and  discussion}

We perform simulations of lattice QCD with 2+1 flavors of rooted staggered 
quarks at imaginary quark chemical potential.
We have made use of the HISQ/tree action~\cite{Follana:2006rc,Bazavov:2010ru} 
as implemented in the publicly available MILC code, 
suitably modified by us in order to introduce an imaginary quark
chemical potential $\mu = \mu_B/3$. 
All simulations make use of the rational hybrid Monte Carlo (RHMC) 
algorithm. The length of each RHMC trajectory has been set to  
$1.0$ in molecular dynamics time units.
Simulations have been done  on lattices of size $16^3\times 6$, {$24^3 \times 6$ and $32^3 \times 8$.}
We have discarded typically not less than one thousand trajectories for each 
run and have collected from 4k to 8k trajectories for each measurement.
The (pseudo)critical  line $\beta_c(\mu^2)$  has been determined as the value for which 
the disconnected susceptibility of the light quark
chiral condensate exhibits a peak.  
To precisely localize the peak, a 
Lorentzian fit has been used. 
Since we want to determine the ratio $T_c(\mu)/T_c(0)$,  we need to set the lattice spacing.
This can be done following the discussion in Appendix B of 
Ref.~\cite{Bazavov:2011nk}, where, for this particular value of $m_l/m_s$, 
the spacing is given in terms of the $r_1$ parameter:
\begin{equation}
\label{scale}
\frac{a}{r_1}(\beta)_{m_l=0.05m_s}=
\frac{c_0 f(\beta)+c_2 (10/\beta) f^3(\beta)}{
1+d_2 (10/\beta) f^2(\beta)} \; ,
\end{equation}
with $c_0=44.06$, $c_2=272102$, $d_2=4281$, $r_1=0.3106(20)\ {\text{fm}}$~\cite{Bazavov:2010hj}
and
\begin{equation}
\label{beta function}
f(\beta)=(b_0 (10/\beta))^{-b_1/(2 b_0^2)} \exp(-\beta/(20 b_0))\;,
\end{equation}
where $b_0$ and $b_1$ are the  universal coefficients of the two-loop beta 
function. 
\begin{figure}[tb]
\centering
\includegraphics*[width=0.6\columnwidth]
{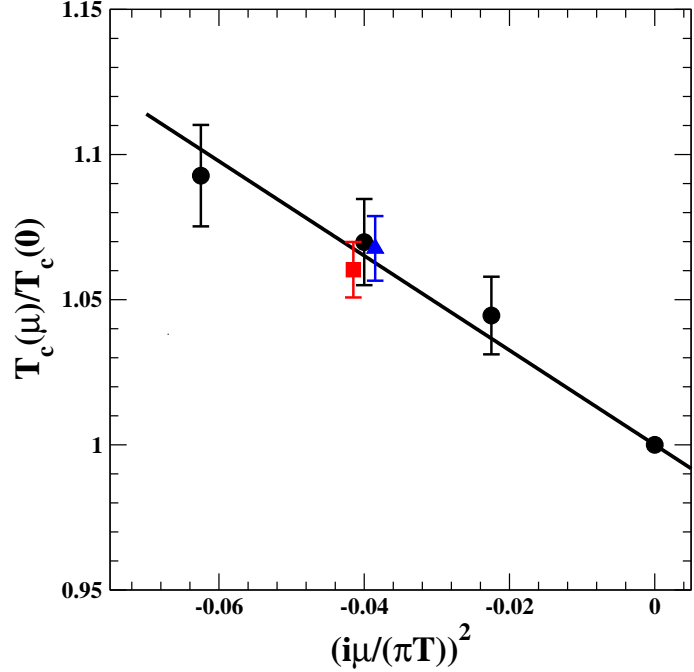}
\caption{$T_c(\mu)/T_c(0)$ versus $((i\mu)/(\pi T))^2$ obtained on a $16^3\times6$ lattice (full circles), on a $24^3\times6$  lattice (full square) and on a $32^3\times8$ lattice (full triangle).
For the sake of readability the abscissae at $((i\mu)/(\pi T))^2=-0.04$  for $24^3\times6$ and $32^3\times8$ data have been slightly shifted.
The full line is a linear fit to the data on the $16^3\times6$ lattice.
}
\label{fig_slope}
\end{figure}
\begin{figure}[t!]
\centering
\includegraphics*[width=0.7\columnwidth]{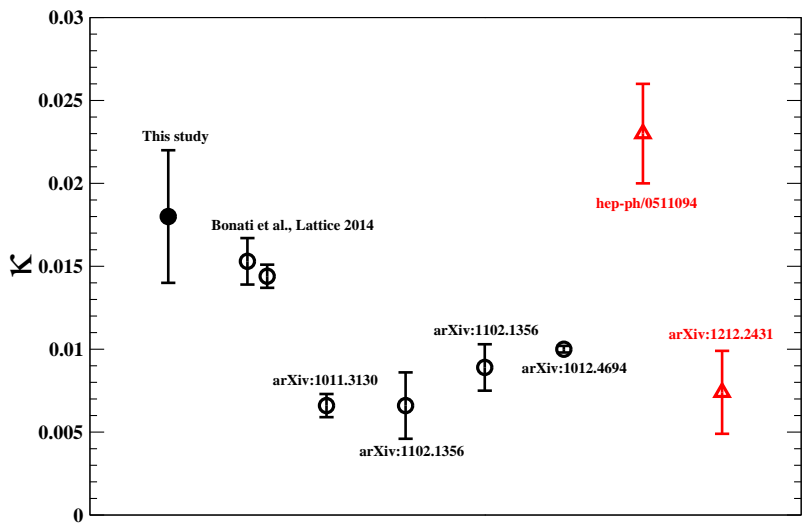}
\caption{Comparison of different determinations of 
the curvature of the chemical freeze-out curve and of the 
(pseudo)critical line for QCD with $n_f = 2+1$.
From left to right: {\em i)} analytic continuation, disconnected chiral 
susceptibility, this study; 
{\em ii)} analytic continuation, renormalized chiral susceptibility, Ref.~\cite{Mesiti:Lattice2014}; 
{\em iii)} analytic continuation, renormalized chiral condensate, Ref.~\cite{Mesiti:Lattice2014}; 
{\em iv)} Taylor expansion, chiral susceptibility, Ref.~\cite{Kaczmarek:2011zz}; 
{\em v)} Taylor expansion, chiral condensate, Ref.~\cite{Endrodi:2011gv}; 
{\em vi)}  Taylor expansion, strange quark number susceptibility, Ref.~\cite{Endrodi:2011gv}; 
{\em vii)} analytic continuation, Polyakov loop, Ref.~\cite{Falcone:2010az}; {\em viii)} freeze-out curvature,
standard analysis, Ref.~\cite{Cleymans:2005xv}; 
{\em ix)} freeze-out curvature, revised analysis of Ref.~\cite{Becattini:2012xb}.
}
\label{k_compare}
\end{figure}
From $a(\beta)$ we determine, for each explored lattice size separately, 
$T_c(\mu)/T_c(0) = a(\beta_c(0))/a(\beta_c(\mu))$. Data for 
$T_c(\mu)/T_c(0)$ versus $\mu/(\pi T)$ are reported in Fig.~\ref{fig_slope}. 
For the $16^3\times6$ lattice, where the determination at three different
values of $\mu$ is available, we have tried a linear fit  in $\mu^2$:
\begin{equation}
\label{linearfit}
\frac{T_c(\mu)}{T_c(0)} = 1 + R_q \left(\frac{i \mu}{\pi T_c(\mu)}\right)^2 \;,
\end{equation}
which works well over the whole explored range ($\chi^2/{\rm d.o.f.} = 0.39$) 
and gives us access to the curvature $R_q$. On the other lattices, assuming 
that linearity in $\mu^2$ still holds, we can extract $R_q$ from the 
determination at $\mu/(\pi T)=0.2i$; we notice that such an assumption, for 
the given value of $\mu$, is consistent with all previous studies on the 
systematics of analytic 
continuation~\cite{Cea:2010md,Cea:2010fh,Cea:2012vi,Cea:2012ev}. 
Our results are:
\begin{eqnarray}
\label{Rq}
R_q(16^3\times6) & =& -1.63(22) \,, \quad \kappa =  0.0183(24) \,, \nonumber \\
R_q(24^3\times6) & =& -1.51(25) \,, \quad \kappa =  0.0170(28) \,,  \\
R_q(32^3\times8) & =& -1.70(29) \,, \quad \kappa =  0.0190(32) \,, \nonumber
\end{eqnarray}
where $\kappa = -R_q/(9 \pi^2)$  is the curvature parameter introduced 
in Eq.~(\ref{curv}).
The results provide evidence that finite size and finite cut-off
systematic effects are within our present statistical uncertainties.
We cannot yet try an extrapolation to the continuum limit of our results,
however, taking into account the statistical errors and the observed variations
of the results with the lattice size and the ultraviolet cutoff, our present
estimate for kappa is
\begin{equation}
\label{curvature}
\kappa = 0.018(4) \;.
\end{equation}
In Figure~\ref{k_compare} we compare our determination of the curvature $\kappa$ defined in Eq.~(\ref{curv}) with other lattice results and
with the estimates of the freeze-out curve. 
A preliminary result from the Pisa group~\cite{Mesiti:Lattice2014}, obtained using the analytic continuation, 
gives $\kappa=0.0153(14)$ from the renormalized chiral susceptibility and
$\kappa=0.0144(7)$  from the renormalized chiral condensate on a $32^2\times8$ lattice.
The
Budapest-Wuppertal collaboration~\cite{Endrodi:2011gv}, using a Symanzik
improved gauge action and stout-link improved staggered fermions on
lattices with temporal size $N_t=6,8,10$ and aspect ratios equal to 
three and four, finds, after continuum extrapolation, 
$\kappa=0.0089(14)$ by the Taylor
expansion method with the strange quark number susceptibility as probe
observable and $\kappa=0.0066(20)$ when, instead, the renormalized chiral 
condensate is used. The Bielefeld-BNL collaboration~\cite{Kaczmarek:2011zz}, 
using the p4-action on lattices with $N_t=4$ and 8, and aspect ratios
up to four, finds $\kappa=0.0066(7)$
again with the Taylor expansion method and the light quark susceptibility as
a probe observable. Another collaboration~\cite{Falcone:2010az} adopted
improved staggered fermions in the p4fat3 version, on lattices with $N_t=4$
and aspect ratio four with physical strange quark mass and pion mass at 220 
MeV, getting $\kappa=0.0100(2)$ by the method of analytic 
continuation, with the Polyakov loop phase as a probe. 
Regarding the freeze-out curve, we report two different estimates. The first
is from the analysis of Ref.~\cite{Cleymans:2005xv}, which is based
on the standard statistical hadronization model; there the authors
parametrize the freeze-out curve as 
\begin{equation}
\label{cleymans}
T_c(\mu_B) = a - b \mu_B^2 - c \mu_B^4\;,
\end{equation}
with $a=0.166(2) \ {\text{GeV}}$, $b=0.139(16) \ {\text{GeV}}^{-1}$, 
and $c=0.053(21) \ {\text{GeV}}^{-3}$, from which we have derived the 
$\kappa$ value reported in Fig.2. 
The second estimate is based on 
the  freeze-out points which are reported in Table I
of Ref.~\cite{Becattini:2012xb} and are based on a modified statistical
reanalysis of the experimental data which includes the effects
of inelastic collisions taking place after freeze-out.
Our result for the curvature is typically between  two and three
standard deviations
larger than previous lattice determinations and seems in a better agreement
with the freeze-out curvature based on the standard statistical hadronization
model.
Possible reasons for the disagreement with previous lattice
determinations can lie in the different methods adopted
to avoid the sign problem, in the different lattice
discretizations, as well as in the different observables
used to locate the transition point, and in the setup of 
quark chemical potentials. 
\begin{figure}[tb]
\centering
\includegraphics*[width=0.6\columnwidth]
{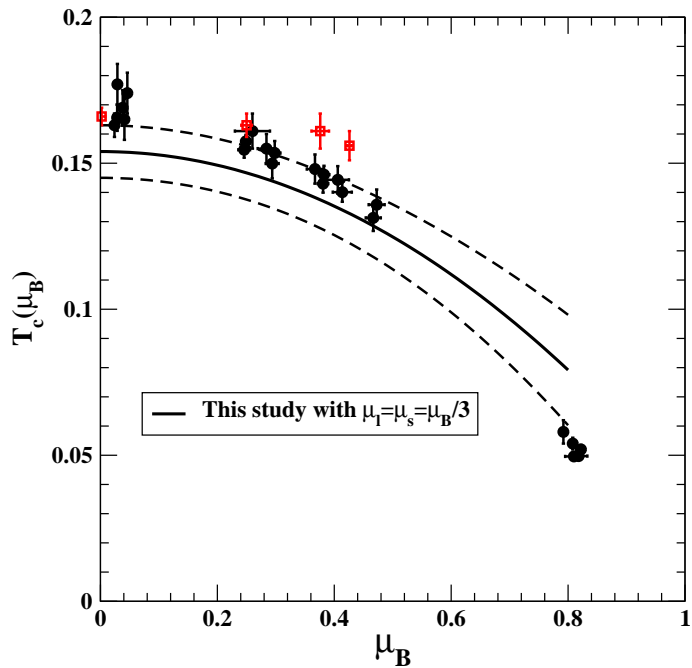}
\caption{$T_c(\mu_B)$ versus $\mu_B$ (units in GeV). Experimental values of $T_c(\mu_B)$ are taken
from Fig.~1 of Ref.~\cite{Cleymans:2005xv} (circles) and from 
Table I of Ref.~\cite{Becattini:2012xb} (squares) 
respectively for the standard and the modified 
statistical hadronization model.
 The solid line is a parametrization corresponding to
$T_c(\mu_B) = T_c(0) - b \mu_B^2$ with $T_c(0)=0.154(9)\, \text{GeV}$ and $b=0.117(27)\,{\text{GeV}}^{-1}$. The dashed lines represent  the corresponding 
error band.}
\label{Tcmu}
\end{figure}
We also report, in Fig.~\ref{Tcmu}, an estimate of the (pseudo)critical line 
which is based on our determination of the curvature. Regarding the value
of $T_c$ at $\mu_B = 0$, which is affected by larger finite 
size and finite cutoff effects than $\kappa$, we refer directly to the 
presently accepted continuum extrapolated value, 
$T_c \sim 155$~\cite{Aoki:2006br,Aoki:2009sc,Borsanyi:2010bp,Bazavov:2011nk},
and in particular to the one obtained in Ref.~\cite{Bazavov:2011nk}
with the same action adopted in our study, 
$T_c(0) = 154(9)$ MeV. From that and from 
$\kappa = 0.018(4)$ we obtain 
$b=0.117(27)\,{\text{GeV}}^{-1}$ (see Eq.~\ref{cleymans}). 
Freeze-out determinations from 
Refs.~\cite{Cleymans:2005xv, Becattini:2012xb} are reported as well.

Let us conclude by discussing the possible sources of systematic effects
in our estimate. One of them is related to the extrapolation from
imaginary to real chemical potentials: in the case of the $16^3\times 6$
lattice we have performed simulations at different values of imaginary
$\mu$, thus verifying that a linear interpolation (in $\mu^2$)
of data works well. For the other two lattices, instead, we have considered
only one value of imaginary $\mu$ ($\mu/(\pi T)=0.2i$) and the linear
behavior has been assumed. Our previous studies based on analytic
continuation, however, indicate that the chosen value of $\mu$ should
lie well inside the region of linearity. Nevertheless,
we plan to perform a more systematic study of this issue.
Finally, we have verified that finite size and cutoff effects are under
control, within the present statistical accuracy. Still, the
extrapolation to the continuum limit, as well as the extension to the physical
value of the light to strange mass ratio, $m_l/m_s \sim 1/28$, and the 
possible effect of varying the strange quark chemical potential, deserve
further investigations and will be the subject of forthcoming works.

\section*{Acknowledgements}

This work was based in part on the MILC Collaboration's public lattice gauge theory code ({\url{http://physics.utah.edu/~detar/milc.html}) and has been partially supported by INFN SUMA project.
Simulations have been performed on BlueGene/Q at CINECA (CINECA-INFN agreement).

\providecommand{\href}[2]{#2}\begingroup\raggedright\endgroup



\begin{thebibliography}{10}

\bibitem{Bazavov:2011nk}
A.~Bazavov, T.~Bhattacharya, M.~Cheng, C.~DeTar, H.~Ding, {\em et.~al.}, {\it
  {The chiral and deconfinement aspects of the QCD transition}},  {\em
  Phys.Rev.} {\bf D85} (2012) 054503,
  [\href{http://xxx.lanl.gov/abs/1111.1710}{{\tt arXiv:1111.1710}}].

\bibitem{Cleymans:2005xv}
J.~Cleymans, H.~Oeschler, K.~Redlich, and S.~Wheaton, {\it {Comparison of
  chemical freeze-out criteria in heavy-ion collisions}},  {\em Phys.Rev.} {\bf
  C73} (2006) 034905, [\href{http://xxx.lanl.gov/abs/hep-ph/0511094}{{\tt
  hep-ph/0511094}}].

\bibitem{Becattini:2012xb}
F.~Becattini, M.~Bleicher, T.~Kollegger, T.~Schuster, J.~Steinheimer, {\em
  et.~al.}, {\it {Hadron Formation in Relativistic Nuclear Collisions and the
  QCD Phase Diagram}},  {\em Phys.Rev.Lett.} {\bf 111} (2013) 082302,
  [\href{http://xxx.lanl.gov/abs/1212.2431}{{\tt arXiv:1212.2431}}].

\bibitem{Philipsen:2005mj}
O.~Philipsen, {\it {The QCD phase diagram at zero and small baryon density}},
  {\em PoS} {\bf LAT2005} (2006) 016,
  [\href{http://xxx.lanl.gov/abs/hep-lat/0510077}{{\tt hep-lat/0510077}}].

\bibitem{Schmidt:2006us}
C.~Schmidt, {\it {Lattice QCD at finite density}},  {\em PoS} {\bf LAT2006}
  (2006) 021, [\href{http://xxx.lanl.gov/abs/hep-lat/0610116}{{\tt
  hep-lat/0610116}}].

\bibitem{deForcrand:2010ys}
P.~de~Forcrand, {\it {Simulating QCD at finite density}},  {\em PoS} {\bf
  LAT2009} (2009) 010, [\href{http://xxx.lanl.gov/abs/1005.0539}{{\tt
  arXiv:1005.0539}}].

\bibitem{Aarts:2013bla}
G.~Aarts, {\it {Complex Langevin dynamics and other approaches at finite
  chemical potential}},  {\em PoS} {\bf LATTICE2012} (2012) 017,
  [\href{http://xxx.lanl.gov/abs/1302.3028}{{\tt arXiv:1302.3028}}].

\bibitem{Follana:2006rc}
E.~Follana {\em et.~al.}, {\it {Highly improved staggered quarks on the
  lattice, with applications to charm physics}},  {\em Phys.Rev.} {\bf D75}
  (2007) 054502, [\href{http://xxx.lanl.gov/abs/hep-lat/0610092}{{\tt
  hep-lat/0610092}}].

\bibitem{Bazavov:2010ru}
A.~Bazavov {\em et.~al.}, {\it {Scaling studies of QCD with the dynamical HISQ
  action}},  {\em Phys.Rev.} {\bf D82} (2010) 074501,
  [\href{http://xxx.lanl.gov/abs/1004.0342}{{\tt arXiv:1004.0342}}].

\bibitem{Cea:2014xva}
P.~Cea, L.~Cosmai, and A.~Papa, {\it {Critical line of 2+1 flavor QCD}},
   {\em Phys.Rev.} {\bf D89} (2014) 074512,
  [\href{http://xxx.lanl.gov/abs/1403.0821}{{\tt arXiv:1403.0821}}].

\bibitem{Bazavov:2010hj}
A.~Bazavov {\em et.~al.}, {\it {Results for light pseudoscalar mesons}},  {\em
  PoS} {\bf LATTICE2010} (2010) 074,
  [\href{http://xxx.lanl.gov/abs/1012.0868}{{\tt arXiv:1012.0868}}].

\bibitem{Mesiti:Lattice2014}
C.~Bonati, M.~D'Elia, M.~Mariti, M.~Mesiti, F.~Negro, and F.~Sanfilippo, {\it
  {The curvature of the QCD critical line from analytic continuation}},  in
  these Proceedings, 2014.

\bibitem{Kaczmarek:2011zz}
O.~Kaczmarek, F.~Karsch, E.~Laermann, C.~Miao, S.~Mukherjee, {\em et.~al.},
  {\it {Phase boundary for the chiral transition in (2+1) -flavor QCD at small
  values of the chemical potential}},  {\em Phys.Rev.} {\bf D83} (2011) 014504,
  [\href{http://xxx.lanl.gov/abs/1011.3130}{{\tt arXiv:1011.3130}}].

\bibitem{Endrodi:2011gv}
G.~Endrodi, Z.~Fodor, S.~Katz, and K.~Szabo, {\it {The QCD phase diagram at
  nonzero quark density}},  {\em JHEP} {\bf 1104} (2011) 001,
  [\href{http://xxx.lanl.gov/abs/1102.1356}{{\tt arXiv:1102.1356}}].

\bibitem{Falcone:2010az}
R.~Falcone, E.~Laermann, and M.~P. Lombardo, {\it {Study of finite temperature
  QCD with 2+1 flavors via Taylor expansion and imaginary chemical potential}},
   {\em PoS} {\bf LATTICE2010} (2010) 183,
  [\href{http://xxx.lanl.gov/abs/1012.4694}{{\tt arXiv:1012.4694}}].

\bibitem{Cea:2010md}
P.~Cea, L.~Cosmai, M.~D'Elia, and A.~Papa, {\it {The phase diagram of QCD with
  four degenerate quarks}},  {\em Phys.Rev.} {\bf D81} (2010) 094502,
  [\href{http://xxx.lanl.gov/abs/1004.0184}{{\tt arXiv:1004.0184}}].

\bibitem{Cea:2010fh}
P.~Cea, L.~Cosmai, M.~D'Elia, and A.~Papa, {\it {The critical line of QCD with
  four degenerate quarks}},  {\em PoS} {\bf LATTICE2010} (2010) 173,
  [\href{http://xxx.lanl.gov/abs/1012.4908}{{\tt arXiv:1012.4908}}].

\bibitem{Cea:2012vi}
P.~Cea, L.~Cosmai, M.~D'Elia, A.~Papa, and F.~Sanfilippo, {\it {Two-flavor QCD
  at finite quark or isospin density}},  {\em PoS} {\bf LATTICE2012} (2012)
  067, [\href{http://xxx.lanl.gov/abs/1210.5896}{{\tt arXiv:1210.5896}}].

\bibitem{Cea:2012ev}
P.~Cea, L.~Cosmai, M.~D'Elia, A.~Papa, and F.~Sanfilippo, {\it {The critical
  line of two-flavor QCD at finite isospin or baryon densities from imaginary
  chemical potentials}},  {\em Phys.Rev.} {\bf D85} (2012) 094512,
  [\href{http://xxx.lanl.gov/abs/1202.5700}{{\tt arXiv:1202.5700}}].

\bibitem{Aoki:2006br}
Y.~Aoki, Z.~Fodor, S.~Katz, and K.~Szabo, {\it {The QCD transition temperature:
  Results with physical masses in the continuum limit}},  {\em Phys.Lett.} {\bf
  B643} (2006) 46--54, [\href{http://xxx.lanl.gov/abs/hep-lat/0609068}{{\tt
  hep-lat/0609068}}].

\bibitem{Aoki:2009sc}
Y.~Aoki, S.~Borsanyi, S.~Durr, Z.~Fodor, S.~D. Katz, {\em et.~al.}, {\it {The
  QCD transition temperature: results with physical masses in the continuum
  limit II.}},  {\em JHEP} {\bf 0906} (2009) 088,
  [\href{http://xxx.lanl.gov/abs/0903.4155}{{\tt arXiv:0903.4155}}].

\bibitem{Borsanyi:2010bp}
S.~Borsanyi {\em et.~al.}, {\it {Is there still any $T_c$ mystery in lattice
  QCD? Results with physical masses in the continuum limit III}},  {\em JHEP}
  {\bf 1009} (2010) 073, [\href{http://xxx.lanl.gov/abs/1005.3508}{{\tt
  arXiv:1005.3508}}].

\end{thebibliography}
\end{document}